\documentclass[runningheads]{llncs}
\usepackage{epsf}
\usepackage{graphicx}
\usepackage{subfigure}
\usepackage{amsmath}
\usepackage{amssymb}       

\usepackage{enumitem}      
   \newenvironment{compenum}{\begin{enumerate}[topsep=0pt,itemsep=0pt,parsep=0pt,partopsep=0pt]}{\end{enumerate}}

\usepackage{color}

\def\half{\mbox{$\frac{1}{2}$}}

\def\x{\mathbf{x}}

\def\b{\mathbf{b}}

\def\u{\mathbf{u}}

\def\0{\mathbf{0}}
\def\1{\mathbf{1}}
\def\l{\mbox{\boldmath $\ell$}}

\def\eps{\epsilon}

\def\Z{{\cal{Z}}}

   \newtheorem{thm}{Theorem}
   \newtheorem{cor}[theorem]{Corollary}

\title{Complexity and approximations for submodular minimization problems on two variables per inequality constraints}

\author{Dorit S. Hochbaum \thanks{Research supported in part by NSF award No. CMMI-1200592.}}
\institute{Department of Industrial Engineering and Operations Research,
         University of California, Berkeley \\ \email{hochbaum@ieor.berkeley.edu}}

\begin{document}
\maketitle

\pagestyle{myheadings} \markright{{\sl }}

\newcommand{\taba}{\hspace*{0.25in}}
\newcommand{\tabb}{\hspace*{.5in}}
\newcommand{\tabc}{\hspace*{.75in}}
\newcommand{\tabd}{\hspace*{1in}}%

\hyphenation{super-optimal}
\hyphenation{approxi-mations}


\begin{abstract}
We demonstrate here that submodular minimization (SM) problems subject to constraints containing up to two variables per inequality, SM2, are $2$-approximable in polynomial time and a better approximation factor cannot be achieved in polynomial time unless NP=P.  When the coefficients of the two variables in each constraint are of opposite signs ({\em monotone} constraints) then the problem of submodular minimization or supermodular maximization is shown here to be polynomial time solvable.  These results hold also for multi-sets that contain elements with integer multiplicity greater than $1$.   Several polynomial time solvable submodular minimization problems are introduced here for the first time, including the submodular closure problem and a submodular cut problem.

Our results indicate that SM2 problems are not much harder than the respective linear integer problems on two variables per constraint.   That is, for monotone constraints both problems are polynomial time solvable, and for all non-monotone NP-hard problems, both problems have $2$-approximation algorithms.  For SM2 problems the factor $2$ approximation is provably best possible, whereas for linear integer problems it has not yet been established that the factor $2$ is best possible, but this has been conjectured. 
On the  other hand, for SM2 problems where the two variables constraints' coefficients form a totally unimodular constraint matrix, the linear integer optimization problem is solved in polynomial time, whereas the submodular optimization is proved here to be NP-hard.

The submodular minimization NP-hard problems for which our general purpose $2$-approximation algorithm applies include: submodular-vertex cover; submodular-2SAT; 
submodular-min satisfiability; submodular-edge deletion for clique submodular-node deletion for biclique and others.  
\end{abstract}

\pagestyle{myheadings} \thispagestyle{plain} \markboth{D. S. Hochbaum
}{Submodular Minimization and Approximations}
\section{Introduction}

We demonstrate here that constrained submodular minimization problems, where each constraint has at most two variables, (SM2), are $2$-approximable in polynomial time.  This approximation factor of $2$ for SM2 is provably best possible unless NP=P.
Furthermore, if the coefficients of the two variables in each constraint have opposite signs, then the submodular minimization problem is solved in (strongly) polynomial time.  All these results extend to multi-sets submodular minimization as well.


A nonnegative function $f$
defined on the subsets of a set $V $ is said to be {\em submodular} if it satisfies
for all $X,Y\subseteq V$, $f(X) + f(Y ) \geq f(X \cap Y ) + f(X \cup Y ).$
A submodular function $f$ is said to be {\em monotone} if
$f(S)\leq f(T)$ for any $S\subseteq T$. 
A binary vector of dimension $n=|V|$, $\x = \{ x_i\}_{i=1}^n$, is associated with a corresponding subset of $V$, $X= \{ i\in V| x_i=1\}$.  The vector $\x$ is then said to be the {\em characteristic vector} of the set $X$.

The SM2 problem addressed here is,

\vspace{-0.1in}
\[
 \hspace{.4in}\begin{array}{ll}
\mbox{~~~~~~~~~~}\  \min \ & f(X) \\
\mbox{(SM2)~~~} \mbox{subject to }\ & a_{ij}x_i +b_{ij} x_j \geq c_{ij} {\mbox { for all  }}  (i,j) \in A\\
\  &   x_j\in \{ 0,1\}  \quad {\mbox { for all  }}  j \in V ,
\end{array}
\]
where $a_{ij}$, $b_{ij}$ and $c_{ij}$ are any real numbers and $A$ is a set of pairs (including singletons, and also allowing multiple copies of the same pair) defining the constraints. Our main results are that any SM2 problem, with constraints that satisfy the {\em round up} property or with monotone submodular objective function, is $2$-approximable in polynomial time and this approximation factor cannot be improved unless NP=P.   A set of constraints satisfies the round up property if any feasible half integer solution can be rounded {\em up} to an integer feasible vector. Vertex cover and all covering constraints satisfy the round up property, but non-covering problems, such as minimum (weighted) node deletion so remaining graph is a maximum clique, satisfy the round up property as well. The formulations and discussion of the properties of these and other SM2 problems is given in Section \ref{sec:examples}.

An inequality constraint in up to two variables, $a_{ij}x_i-b_{ij} x_j \geq c_{ij}$ is called {\em monotone} if $a_{ij}$ and $b _{ij}$ have the same signs. (This concept of monotonicity is unrelated to the monotonicity of a submodular function.) The problem of submodular minimization or supermodular maximization on monotone constraints is shown here to be polynomial time solvable.  This is in stark contrast to submodular minimization or supermodular maximization over constraints with totally unimodular constraints matrix which is proved here to be NP-hard.  This demonstrates that monotone constraints form a more significant structure than totally unimodular constraints, in terms of complexity, for submodular (supermodular) minimization (maximization).


The results here all apply to submodular minimization on {\em multi-sets}, (SM2-multi). These are submodular functions defined on sets containing
elements with multiplicity greater than $1$.  A nonnegative integer
vector $\x \in \Z ^n$ is the characteristic vector of a multiset $X=\{
(i,q_i)| x_i=q_i\}$, where $(i,q_i) \in X$ means that $X$ contains element $i$ $q_i$ times, for positive integers $q_i$. All properties of submodular functions extend to multi-sets, with the generalized
definition of containment, $X_1\subseteq X_2$ to mean that for all
$(i,q_i)\in X_1$, $(i,q'_i)\in X_2$ with $q_i\leq q'_i$.  The problem of constrained submodular minimization on multi-sets is $\min \{ f(X)|A\x \ge \b,\ \0 \leq \x \leq \u , \x \in \Z^n \}$. Let the upper bound on the multiplicity of element $i$ be $u_i$.  The formulation of SM2-multi is then,

\vspace{-0.01in}
\[
 \hspace{.4in}\begin{array}{ll}
\mbox{~~~~~~~~~~~~\ \ \ }\ \  \min \ & f(X) \\
\mbox{(SM2-multi)}\  \mbox{subject to }\ & a_{ij}x_i +b_{ij} x_j \geq c_{ij} {\mbox { for all  }}  (i,j) \in A\\
\  &   0\leq x_j \leq u_j  \quad {\mbox { for all  }}  j \in V ,
\end{array}
\]

The respective $2$-approximations or polynomial time algorithms for multi-sets are attained in time polynomial in $U=\max _{j=1,\ldots n} u_j$.  The dependence of the run time on $U$ cannot be removed (to, say, logarithmic dependence) unless NP=P.

\subsection{Related research}
A prominent example of SM2 is the submodular vertex cover, SM-vertex cover, where the constraint matrix $A$ contains exactly two $1$s per row and $\b$ is a vector of $1$s. 

Approximating SM-vertex cover has been a subject of previous research work.  Three different $2$-approximation algorithms were devised for the problem:  Koufogiannakis and Young \cite{KY09} devised approximations for SM- ``covering" problems with monotone submodular objective function.  The approximation algorithm is based on the frequency technique (called maximal dual feasible technique in \cite{Hoc97} Ch. 3).
Their algorithm is a $2$-approximation for the SM-vertex cover for monotone submodular objective function.  Goel et al.\ \cite{GKTW} devised a $2$-approximation algorithm for SM-vertex cover with monotone submodular function which involves solving a relaxation with the Ellipsoid method with a separation algorithm equivalent to a submodular minimization problem.  Goel et al.\ further proved that submodular vertex cover is inapproximable within a factor better than $2$.   Iwata and Nagano in \cite{IN} presented a $2$-approximation algorithm for the SM-vertex cover, and addressed the SM-set cover and the SM-edge cover. Their algorithm does not require the submodular function to be monotone. Iwata and Nagano's technique relies on using Lov\'{a}sz extension of submodular minimization to convex minimization.

\subsection{Contributions here}
We devise here a unified framework for generating $2$-approximation algorithms for all NP-hard SM2 and SM2-multi problems with constraints that have the round-up property (that include all covering matrices), or, if round-up does not hold, for monotone submodular functions.  Unlike previous results, these algorithms do not require solving a linear programming relaxation or using the Lov\'{a}sz extension convex optimization, yet run in strongly polynomial time (Theorem \ref{thm:approx}).  In particular, our algorithm is a $2$-approximation algorithm for the SM-vertex cover (without restriction of submodular function's monotonicity).  Other NP-hard submodular
minimization problems for which we derive $2$-approximations include: The submodular
min-2SAT; minimum node deletion biclique; minimum edge deletion clique; and min SAT.  Among these only the SM-min-2SAT requires monotone submodular objective function.  

In addition to approximation algorithms we provide polynomial time algorithms for SM2 over {\em monotone constraints} which are of the form $a_{ij}x_i -b_{ij} x_j \geq c_{ij}$, where $a_{ij}$ and $b_{ij}$ are of the same signs.  Such problems include the submodular $s,t$-cut problem and the closely related submodular closure problem.  The latter problem is defined for a precedence relationship formalized as a directed graph, $G=(V,A)$.  A subset $S \subseteq V$ is said to be closed if it contains all the successors (or predecessors) of $S$.  The SM-closure problem is to find a closed set that maximizes (or minimizes) a supermodular (or submodular) objective function.   A corollary of this result, is that the minimum bi-submodular vertex cover, and maximum bi-supermodular independent set, in bipartite graphs are solved in polynomial time, yet the submodular or supermodular optimization over bipartite graphs, is NP-hard.  (Bi-submodular and bi-supermodular functions are defined in Section \ref{sec:notation}. The NP-hardness of SM-vertex cover and SM-independent set on bipartite graphs and totally unimodular matrices is proved in Section \ref{sec:VCbip}.)

Our results shed some light on the relationship between submodular minimization and linear minimization in integers.   Obviously submodular minimization can only be harder than integer minimization.  Yet, on the one hand our results imply that for two variables per inequality constraints, submodular minimization is not harder than the respective integer linear optimization.  Indeed, for any NP-hard integer program on two variables per inequality constraints, Hochbaum et al. devised a unified $2$-approximation algorithm \cite{HMNT}, which the algorithm here generalizes. For the linear vertex cover case,  that algorithm transforms the problem to vertex cover on bipartite graph, which is solvable in polynomial time.  On the other hand, as we show in Section \ref{sec:VCbip} Theorem \ref{thm:SM2-TU-hard}, SM-vertex cover on bipartite graphs is an NP-hard problem, whereas the linear vertex cover on bipartite graphs in polynomial time solvable.  This demonstrates that submodular minimization over a totally unimodular constraint matrix, is NP-hard, and in that sense strictly harder than the respective integer linear optimization.

\noindent
{\bf{Summary of contributions:}}
\begin{compenum}
\item We present here the first known polynomial time $2$-approximation algorithms for a large
family of constrained submodular minimization problems, SM2 and SM2-multi, with constraints that contain at most
two variables per inequality.
\item  SM2 problems on monotone constraints are shown to be solved in strongly polynomial time for either submodular minimization or supermodular maximization.  This holds also for multi-sets, with complexity that depends on the multiplicity of the sets. This complexity cannot be improved, as the linear version of SM2 on monotone constraints is (weakly) NP-hard.
\item Submodular minimization over constraints with coefficients' matrix that is totally unimodular is proved to be NP-hard in Theorem \ref{thm:SM2-TU-hard} .  In particular, submodular vertex cover on bipartite graphs is an NP-hard problem.  This proof provides additional evidence to the difficulty of generalizing linear optimization, or approximation, algorithms to the submodular context.
\item Iwata and Nagano \cite{IN} use a construction based on Lov\'{a}sz convex extension of submodular functions in order to prove the validity of their $2$-approximation algorithm for SM2-vertex cover.  This construction involves an ``intermediate" convex formulation that is shown here to be unnecessary.  We provide a direct and simple proof of the $2$-approximability using only the properties of submodular functions.
\item The $2$-approximation factor is best possible approximation factor for
all SM2 problems.  This is shown to follow from the lower bound proof on the approximability of SM-vertex cover of  Goel et al.\ \cite{GKTW} which establishes that SM-vertex cover cannot be approximated in polynomial time within a factor of $2-\eps$, for any $\eps >0$, unless $NP=P$.
\item As a special case of submodular minimization over totally unimodular constraints, the submodular vertex cover on bipartite graph is shown here to be an NP-hard problem.  But the bi-submodular vertex cover, and bi-supermodular independent set, on bipartite graphs, are polynomial time solvable.
\end{compenum}

\section{Notations and preliminaries}\label{sec:notation}
Given an ${m\times n}$ real matrix $A^{(2)}$  where each row contains at most two
non-zeroes, the SM2 problem can be written as $\min \{ f(X) |A^{(2)}\x \ge \b,\ \0\leq  \x \leq \u , {\mbox {integer}} \}$ for $u_j=1$ for $j=1,2,\ldots ,n$. For general positive values of $u_j$s the problem is called SM2-multi.

An important class of SM2 has the two non-zeroes in each row of $A^{(2)}$ of opposite signs in which case the constraints are said to be monotone.  SM2 problems on monotone constraints are shown here to be polynomial time solvable.

A feasible SM2 (that has a feasible integer solution) is said to have the {\em round-up} property, if for any given feasible half integral solution vector $\x ^{\half}$ there exists an integer feasible solution $\x ^{\rm int}$ such that $\x ^{\half}\leq \x ^{\rm int}$.  We refer to an SM2 with such set of constraints as {\em round-up-SM2}.  Notice that the round-up property depends on the constraints only.   All covering matrices, where the inequalities are $\geq$ constraints and all coefficients are non-negative have the round-up property.  But there are also non-covering matrices that have the round-up property.

For a directed graph $G=(V,A)$ and $B,D\subseteq V$, we denote by $(B,D)$ the
set of arcs from nodes in $B$ to nodes in $D$, $(B,D)=\{ (i,j)|i\in B, j\in
D\}$. Note that $B$ and $D$ need not be disjoint and can be equal. For a bi-partition, we refer to the set of arcs in  $(B,\bar{B})$ as the {\em cut-set}. In
$G=(V,A)$ a set of nodes $D\subseteq V$ is said to be {\em closed} if all the
successors (or predecessors) of the nodes in $D$ are also in $D$.  In other words, the {\em
transitive closure} of $D$, forming all the nodes reachable from nodes of $D$
along a directed path in $G$, is equal to $D$.  We call linear constraints on two variables of the form $x\leq y$ {\em closure constraints}.  Closure constraints are obviously monotone.

A directed graph is said to be a {\em closure graph} if all arcs are of infinite capacity.
An $s,t$-graph $G_{st}=(V\cup \{ s,t\},A\cup A_s\cup A_t) $ with
$A_s$ the set of arcs adjacent to source $s$ and $A_t$ a set of arcs adjacent
to sink $t$, is called a {\em closure $s,t$-graph} if all finite capacity arcs are either in $A_s$ or $A_t$.

A function $f$ is said to be {\em supermodular} if for all $X,Y\subseteq V$,
$f(X) + f(Y ) \leq f(X \cap Y ) + f(X \cup Y ).$

A function $f$ defined on a bipartite graph $G=(V_1\cup V_2,E)$ is {\em bi-supermodular} if there exist two supermodular functions $f_1$ and $f_2$ defined on $V_1$ and $V_2$ respectively, such that, for $X_1=X\cap V_1$, $X_2=X\cap V_2$,
 $$f(X)=f_1(X_1)+f_2(X_2).$$
 A bi-submodular function is defined analogously.

\section{Some of the submodular minimization problems solved here}\label{sec:table-examples}

Table \ref{tab:1} lists several SM2 submodular problems for which the algorithmic framework devised here applies.  In the table it is noted, for each problem, whether it has the round-up property or not.  Problems that are polynomial time solvable with the technique here are indicated with an approximation factor of $1$.

\begin{table} [htb]
\begin{center} 
\begin{tabular}{|l|l|l|l|l|} \hline
{\bf SM-Problem  } & {\bf Monotone } & {\bf  Round-up}  & {\bf Submodular}& {\bf Aprrox}\\
{\bf Name} & {\bf Constraints} & {\bf Property}  & {\bf Objective $f()$} & {\bf Factor}
\\ \hline \hline
Vertex cover & No & Yes & any &2\\
Complement of max-clique & No & Yes  &  any&2\\
Node-deletion bi-clique & No & Yes  & any&2\\
Min-satisfiability &  No & Yes  & any &2\\
Min-2SAT &  No & No & monotone&2\\
SM-closure &  Yes & NA & any&1\\
SM-cut  &  Yes & NA & any&1\\
\hline \hline
\end{tabular}
\vspace{0.1in}\caption{Examples of $2$-approximable and polynomial time solvable SM-problems. } \label{tab:1}
\end{center}
\end{table}

We now provide the formulations and discussion of properties for each of the problems. SM-closure and SM-cut are discussed in Section \ref{sec:closure}.

\subsection{Formulations of several SM2 problems}\label{sec:examples}

{\bf Vertex cover}.  The vertex cover problem is to find a subset of nodes in a
graph $G=(V,E)$ so that each edge in $E$ has at least one endpoint in the
subset.

\[
({\mbox{\sf SM-vertex-cover}}) \hspace{.1in}\begin{array}{ll}
\mbox{} \min \ & f(X)\\
\mbox{subject to }\ & x_i+x_j \geq 1 \ \
\mbox {for all} \quad  [i,j]  \in E \\
&   x_i  \ \
\mbox {binary} \quad   i\in V.
\end{array}
\]
The submodular vertex cover problem was shown to have a
$2$-approximation by Iwata and Nagano, \cite{IN} for general submodular $f()$.   SM-vertex cover obviously has the round-up property and therefore the $2$ approximation described here applies to any general submodular objective function.  When the graph $G=(V_1\cup V_2,E)$ is bipartite, the SM-vertex-cover is still NP-hard (Section \ref{sec:VCbip}), but for a bi-submodular objective, $f(X_1\cup X_2))=f_1(X_1)+f_2(X_2)$, for $X_i\subseteq V_i$, $i=1,2$, the problem is polynomial time solvable.

{\bf Complement of maximum clique}.  The maximum clique problem is a well known
optimization problem that is notoriously hard to approximate, e.g.\
H{a}stad, \cite{Ha96}. The problem is to find in a graph the largest set of
nodes that forms a clique -- a complete subgraph.

An equivalent statement of the clique problem is to find the complete subgraph
which maximizes the number (or more generally, sum of weights) of the edges in
the subgraph. When the weight of each edge is $1$, then there is a clique of
size $k$ if and only if there is a clique on $k \choose 2$ edges. The
inapproximability result for the node version extends trivially to this edge
version as well.

The complement of this edge variant of the maximum clique problem is to find a
minimum weight of edges to {\em delete} so the remaining subgraph induced on
the non-isolated nodes is a clique.  We define here the SM-edge deletion for
clique.   For a graph $G=(V,E)$,  the submodular function $f(Z)$ is defined on
the set of variables $z_{ij}$ for all edges $[i,j]\in E$.  Let $x_j$ be a
variable that is $1$ if node $j$ is in the clique, and $0$ otherwise.  Let
$z_{ij}$ be $1$ if edge $[i,j]\in E$ is {\em deleted}.
\[
({\mbox{\sf SM-Clique-edge-delete}}) \hspace{.01in}\begin{array}{ll}
\mbox{} \min \ & f(Z)\\
\mbox{subject to }\ & 1- x_i\leq z_{ij} \quad   [i,j] \in E\\
 & 1- x_j\leq z_{ij} \quad   [i,j] \in E\\
 & x_i+x_j\leq 1 \quad   [i,j] \notin E\\
 &    x_j  \ \
\mbox {binary} \quad   j \in V \\
&    z_{ij}  \ \
\mbox {binary} \quad   [i,j] \in E.
\end{array}
\]

This formulation has two variables per inequality and therefore it is SM2 and a
$2$-approximation algorithm exists. The gadget and network for solving the
monotonized {\sf SM-Clique-edge-delete}
problem are given in detail in \cite{Hoc02}.   This SM2 is of covering-type and therefore the approximation algorithm applies to any submodular function.

{\bf Node deletion biclique}. Here we consider the submodular minimization of
node deletion in a bipartite graph $(V_1\cup V_2,E)$ so that the subgraph induced on the remaining nodes forms a biclique (a complete bipartite graph).  This problem is identical to the submodular vertex
cover on a bipartite graph, proved in Section \ref{sec:VCbip} to be NP-hard.  The linear version of this problem is polynomial time solvable, \cite{Hoc98}.
In the formulation given below $x_i$ assumes the value $1$ if node $i$ is
deleted from the bipartite graph, and $0$ otherwise.

\[
({\mbox{\sf SM-Biclique-node-delete}}) \hspace{.01in}\begin{array}{ll}
\mbox{} \min \ & f(X)\\
\mbox{subject to }\ &
  x_i+x_j \geq 1  \
\mbox{for edge $\{ i,j\}\not \in E$~~$i\in V_1, j\in V_2$} \\
  &  x_j \in \{0,1\} \ \ \mbox{for all~}j \in V_1\cup V_2.
\end{array}
\]

{\bf Minimum satisfiability}. In the problem of {\em minimum satisfiability},
MINSAT, we are given a CNF satisfiability formula.  The aim is to find an
assignment satisfying the smallest number of clauses, or the smallest weight
collection of clauses. The MINSAT problem was introduced by Kohli et.\ al.\
\cite{KKM94} and was further studied by Marathe and Ravi \cite{MR96}.  The
problem is NP-hard even if there are only two variables per clause, \cite{KKM94}.

The submodular minimum satisfiability SM-MINSAT problem can be
formulated as SM2, and thus $2$-approximable: Choose a binary variable $y_j$
for each clause $C_j$ and a binary variable $x_i$ for each literal.  Let $S^+(j)$ be the set of
variables that appear unnegated and $S^-(j)$ those that are negated in clause
$C_j$.  The following formulation of MINSAT has two variables per inequality
and is thus a special case of SM2:
\[
({\mbox{SM-MINSAT}}) \hspace{.2in}\begin{array}{ll}
\mbox{} \min \ & f(Y)\\
\mbox{subject to }\ & y_j \geq x_i \quad
\mbox{~~~~~for $i\in S^+(j)$} \ \
\mbox{for clause $C_j$} \\
  &  y_j \geq 1- x_i \quad
\mbox{for $i\in S^-(j)$} \ \
\mbox{for clause $C_j$} \\
  &  x_i, y_j \ \ \mbox{binary for all~} i,j.
\end{array}
\]
It is interesting to note that the formulation is monotone when for all clauses
$C_j$ $S^+(j)=\emptyset$ or in all clauses $S^-(j)=\emptyset$. (In the latter
case need to transform the $x$ variables to $x'$ with $x'=-x$.)  Indeed in
these instances the boolean expression is uniform and the problem is trivially
solved setting all variables to FALSE in the first case, or to TRUE in the
latter case.

Although SM-MINSAT is not of covering type, it is nevertheless a round-up SM2, as can be easily verified, and therefore there is no restriction on $f()$ for the $2$-approximation algorithm to apply.

{\bf MIN-2SAT}.  The MIN-2SAT problem is defined for a 2SAT CNF with each
clause containing at most two variables.  The goal is to find a truth assignment, satisfying all clauses, with the least weight
collection of variables that are set to true.  Although finding a satisfying assignment to a 2SAT can be done
in polynomial time, Even et al.  \cite{EIS}, finding an assignment that
{\em minimizes} the number, or the weight, of the true variables in NP-hard.

Let $X$ be the set of true variables, and $x_i=1$ if the $i$th variable is set
to true, and $0$ otherwise. \vspace{-0.1in}
\[
({\mbox{SM-MIN-2SAT}}) \hspace{.2in}\begin{array}{ll}
\mbox{~} \min \ & f(X)\\
\mbox{subject to }\ & x_i+x_j\geq 1 \quad

\mbox{~~for clause $(x_i\vee x_j)$} \\
& x_i-x_j\geq 0 \quad
\mbox{~~for clause $(x_i\vee \bar{x_j})$} \\
& x_i+x_j\leq 1 \quad
\mbox{for clause $(\bar{x_i}\vee \bar{x_j})$} \\
 &  x_i \ \ \mbox{binary for all~} i=1,\ldots ,n.
\end{array}
\]

Each constraint here has up to two variables and thus this problem is in the
class SM2. Consequently we get for this problem a $2$-approximation in polynomial time.
The general SM-MIN-2SAT does {\em not} have the round-up property and therefore the $2$-approximation algorithm applies for $f()$ {\em monotone} submodular function.

Additional problems related to finding maximum biclique -- a clique in a
bipartite graph -- are also formulated in two variables per inequality, in
\cite{Hoc98}.  For these problems, all the corresponding submodular minimization problems
are either monotone, and thus solved in polynomial time, or have a polynomial
time $2$-approximation.

\section{The submodular closure and submodular cut problems}\label{sec:closure}

\subsection{The submodular closure problem}
The submodular closure problem is defined on a directed graph $G=(V,A)$.  A set $S \subseteq V$ is closed if it contains the transitive closure of $S$.
The {\em SM-closure} problem is to find a closed set in the graph that minimizes a submodular function.
The submodular optimal closure is a generalization of the (linear) closure problem
defined on a directed graph $G=(V,A)$.   Consider first the maximum weight closure problem where the closure is in terms of successors. Let $x_j$ be a binary variable that is $1$ if node $j$ is in the closure, and $0$ otherwise. Let $w_j$ be the weight of node $j$.  Note that the problem is trivial if all $w_j$ are positive (optimal solution is $V$), or if all $w_j$ are negative (optimal solution is $\emptyset$). The problem formulation is,
\[
 \hspace{.4in}\begin{array}{ll}
 \mbox{(max-closure)~~}  \max \ &
\sum_{j \in V} w_j x_j \\
~~~~~ \mbox{subject to }\ & x_i\leq x_j \quad \forall (i,j)\in A, \\ &    x_j
\ \ \mbox {binary} \quad   j \in V.
\end{array}
\]

To solve this linear problem we set an $s,t$ closure graph: Consider the partition on $V$ to $V^+=\{ v\in V| w_v>0\}$ and $V^-=\{ v\in V| w_v\leq 0\}$, and note that both sets are non-empty for non-trivial problems.  We add to the graph a node $s$ and a set of arcs $\{(s,j)| j\in V^+\}$, where arc $(s,j)$ is of capacity $w_j$.  Next we add a node $t$ and a set of arcs $\{(i,t)| i\in V^-\}$, where arc $(i,t)$ is of capacity $-w_i$.  All arcs in $A$ are assigned infinite capacity.  It is easy to see that the source set $S$ for any finite $s,t$-cut $(S,T)$ in this graph is a closed set, and that the minimum capacity cut has a source set of maximum weight.  Additional details on the closure problem can be found, e.g.\ in \cite{HQ03}.

The minimum closure problem can be solved by either replacing the weights by their negation, or the optimal solution is the sink set $T$ of a minimum cut partition $(S,T)$, that is closed with respect to predecessors.

%
%
%
A non-binary integer version of the minimum closure problem with a convex
separable objective replacing the linear objective was shown to be solvable in
polynomial time by a parametric cut algorithm in Hochbaum and Queyranne
\cite{HQ03}.

Now consider the SM-closure, for a submodular function $f(X)$ where $X \subseteq V$:

\[
 \hspace{.4in}\begin{array}{ll}
 \mbox{(SM-closure)~~}  \min \ &
 f(X) \\
~~~~~ \mbox{subject to }\ & x_i\leq x_j \quad \forall (i,j)\in A, \\ &    x_j
\ \ \mbox {binary} \quad   j \in V .
\end{array}
\]
Note that this problem is interesting, or non-trivial, for non-monotone submodular functions.
The algorithm for solving the SM-closure problem is the key subroutine in solving all the problems
presented here.  To solve this problem one can use any submodular minimization over a ring:
The collection of closed sets form a {\em
ring} since their union and intersection are also closed sets.
Submodular minimization over a ring family was first shown
to be solved in strongly polynomial time by Gr\"{o}tschel, Lov\'{a}sz,
and Schrijver in \cite{GLS}.  Combinatorial strongly polynomial algorithms
were given later by Schrijver, \cite{Sch00} and by Iwata, Fleischer, and
Fujishige \cite{IFF}. The current fastest strongly polynomial algorithms on a
ring family were devised by Orlin \cite{Orl}, and later by Iwata and Orlin,
\cite{IOrlin}.

\subsection{The submodular cut problem}
The SM-cut problem is closely related to SM-closure in an analogous manner to the respective relationship between minimum $s,t$-cut and linear closure problems.  Let SM-cut be defined on an $s,t$ graph $G_{st}$.  For $S\subseteq V$, the source set of the cut, the set of arcs from $\{ s\}\cup S$ to $\bar{S}\cup \{ t\}$, also denoted as $(\{ s\}\cup S,\bar{S}\cup \{ t\})$, form the cut-set.  The problem of finding such partition so that the submodular function value  $f(S)$ of the associated cut-set is minimum, is solvable in polynomial time.  To see that, observe that the source sets $S$ form a ring: the intersection and union of source sets of cuts are source sets of cuts.

We are interested here in a special case of the submodular cut problem defined on an
$s,t$ {\em closure} graph $G_{st}$ that has all finite capacity arcs adjacent to source or to sink.  The cut capacity is a submodular function defined on subsets of arcs adjacent to source and sink, in $A_s\cup A_t$.
For $S\subseteq V$ and the associated bipartition $(\{ s\}\cup S,\bar{S}\cup \{ t\})$, the set of arcs in the cut is $\{ A_s \cap (\{ s \} , \bar{S})  \cup (A_t \cap (S,\{ t\}))\} .$  The restriction that the graph is a closure graph means that no arc in $A$, that does not have one endpoint equal to $s$ or $t$, can be in the cut-set.  Therefore any finite submodular capacity cut must have the source set closed in the associated graph $G$.  It follows that the submodular cut on closure graphs problem is equal to the SM-closure problem.

\section{$2$-approximation for SM2 and SM2-multi}\label{sec:approx}
\subsection{A polynomial time algorithm for solving SM2 and SM2-multi on monotone constraints}\label{sec:binarizing}
We use here a process, first introduced by Hochbaum and Naor \cite{HN}, that maps a monotone constraint in integers $a_{ij}x_i-b_{ij}x_j \geq c_{ij}$ into an equivalent collection of closure constraints on binary variables.  That is, each constraint is of the form $x' \leq y'$ for $x'$, $y'$ binary variables.

A sketch of the procedure is as follows: We first replace the integer variables by binary variables.  This step in not required for SM2 where the variables are already binary.  For each element $i$ and $p=1,\ldots ,u_i$ we let $x_i^{(p)}=1$ only if $x_i \geq p$.  In particular the variable $x_i$ can be written as a summation of
binary variables $x_i=\sum_{p=1}^{u_i} x_i^{(p)}$.  The following restriction then holds for all $p=1,\dots,u_i$:  $x_i^{(p)}=1 \implies
x_i^{(p-1)}=1.$  

For a monotone constraint $a_{ij}x_i-b_{ij}x_j \geq c_{ij}$, let $q(p)\equiv \lceil{\frac{c_{ij}+b_{ij}p}{a_{ij}}}\rceil$.
Then this monotone constraint can be equivalently written as follows:
\[
 \hspace{.4in}\begin{array}{ll}
 x_j^{(p)} \le x_i^{(q(p))} & \quad\forall\ (i,j)\in A\quad\text{for }p=1,\dots,u_j\\
 x_i^{(p)}\le x_i^{(p-1)} & \quad\forall\ i\in V\quad\text{for }p=1,\dots,u_i \\
 x_i^{(p)}\in \{ 0,1\} & \quad\forall\ i\in V\quad\text{for }p=1,\dots,u_i.
\end{array}
\]

Note that for binary variables $x_j^{(0)}=1$ and there is only one inequality generated for $x_j^{(1)}$.  If $q(p)>1$ then $x_j$ is fixed at $0$ and removed from the set of variables; if $q(p)<1$ then the constraint is trivially satisfied and may be removed.

%

These constraints are closure constraints in an $s,t$-graph where for each element $i$ there are $u_i$ nodes and for every constraint $(i,j)\in A$ in the original problem there are up to $\min \{u_i,u_j\}$ arcs.
Hence any SM2 or SM2-multi problem on monotone constraints is equivalent to a SM-closure
problem on a graph with $O(\sum _{i\in V} u_i)$ nodes, and is solvable in polynomial time in the size of the graph.  For SM2-multi if the range of the variables is not a polynomial quantity then the size of the graph is {\em pseudopolynomial}.  As noted previously, this run time cannot be improved to
polynomial time unless NP=P, since finding a feasible solution to a monotone integer program on
constraints with up to two variables per inequality is NP-hard,
\cite{Lagarias}.  (The pseudopolynomial run time of the algorithm for monotone
constraints in \cite{HN} indicates that the problem is in fact {\em weakly} NP-hard.)

\subsection{Transforming general SM2 and SM2-multi to their monotonized version}\label{sec:monotonize}
General SM2s (and SM2-multi) are transformed to
monotone SM2s using a process we refer to as {\em monotonizing}.  Monotonizing changes the objective function as well.  The monotonized version of an SM2-multi is a relaxation of the problem that is solved in polynomial time as shown in Section \ref{sec:binarizing}.

The monotonizing process is described here for general SM2 and SM2-multi.  We first duplicate the set of elements $V$ to $V^+$ and $V^-$, and the characteristic vectors $\x$ to $\x ^+$ and $\x ^-$, so that $x_j^+$ assumes values in $\{ 0, 1,\ldots ,u_j\}$ and $x_j^-$ assumes values in $\{ -u_j, \ldots ,-1,0\}$.  If $x_j^- =-p$ then the corresponding set in $V^-$ contains element $j$ with multiplicity $p$.  $\x ^+$ is the characteristic vector of subsets of $V^+$ and $\x'=-\x ^-$ is the characteristic vector of subsets of $V^-$.

Each non-monotone inequality $a_{ij}x_i + b_{ij}x_j \geq c_{ij}$ is replaced by the
following two inequalities:
\vspace{-0.021in}
\begin{align*}
a_{ij} x^+_i - b_{ij} x^-_j \geq c_{ij}\\
-a_{ij} x^-_i + b_{ij} x^+_j \geq c_{ij},\\
\end{align*}

\vspace{-0.265in}

and in case the non-monotone constraints contain a subset of monotone constraints, each monotone inequality $a'_{ij}x_i - b'_{ij}x_j \geq c'_{ij}$ is replaced by the two inequalities:
\vspace{-0.021in}
\begin{align*}
a'_{ij} x^+_i - b'_{ij} x^+_j \geq c_{ij}\\
-a'_{ij} x^-_i + b'_{ij} x^-_j \geq c_{ij}.\\
\end{align*}

\vspace{-0.221in}

It is easy to verify that setting $x_j=\frac{x_j^+ -x_j^-}{2}$ is feasible for the original inequalities, and $x_j=x_j^+ -x_j^- $
is feasible to the original inequalities multiplied by $2$,  $2a_{ij}x_i + 2b_{ij}x_j \geq 2c_{ij}$.  We refer to the latter as the {\em doubled} inequalities.

\subsection{The $2$-approximation algorithm}
For any given SM2-multi the monotonized problem is solvable with the polynomial time algorithm described in Section \ref{sec:binarizing}.    We refer to the resulting SM-closure problem as {\em relaxed} SM2 (or SM2-multi).  The relaxed SM2-multi problem is defined on binary variables that form characteristic vectors of subsets of $V^+$ and $V^-$.  These sets of elements are represented as nodes, and every inequality of the form $ y' \leq z'$ corresponds to an arc $(y',z')$ in the set of arcs $A'$.  This SM closure is then defined on the graph  $G=(V^+\cup V^-,A')$.

With the substitution, $ x'_i= -x^-_i \ \mbox{ for all    }\   i\in V^-$, for any closed set $S$, $V^+\cap S=\{ j\in V^+|x_j^+=1\}$ and $V^-\cap S= \{ j\in V^-|x_j'=0\} $.
We denote $X^+=\{ j\in V^+|x_j^+=1\}$ and $X^-=\{ j\in V^-|x_j'=1\} $.  That is, the vectors $\x ^+$ and $\x '$ are the characteristic vectors of the sets $X^+$ and $X^-$.

Let $S^{*}$ be an
optimal set minimizing the function $f()$ for the (original) SM2 formulation, and let $\x^*$ be the associated characteristic vector.  On the graph $G=(V^+\cup V^-,A')$, let $S^{*+}$ and $S^{*-}$ be the copies of $S^{*}$ in $V^+$ and $V^-$ respectively.  This is a feasible solution for the relaxed problem since $S^{*+}\cup (V^- \setminus S^{*-})$ is a closed set, and the vectors $\x^{*+}=\x'^{*}=\x^*$ defined on $V^+$  and $V^-$ are the characteristic vectors of $S^{*+}$ and $S^{*-}$, and the setting $x_i={x^{*+}_i+x'^{*}_i}$ is a feasible solution $\x$ for the ``doubled" inequality constraints.

The submodular function $f()$ is defined on subsets of $V$ and therefore the objective function of the SM-closure problem defined on the constructed graph, $g(X^+\cup X^-)= f(X^+) + f(X^-)$, is well defined.  It is easy to prove that if $f()$ is a submodular function then $f^+(D)=f(V^+\cap D)$ and $f^-(D)=f(V^-\setminus D)$ are submodular functions. Therefore, $f(X^+)$, $f(X^-)$ and $g(X^+\cup X^-)$ are submodular functions.



\begin{thm}\label{thm:approx}
Let $X'^+\subseteq V^+$ and $X'^-\subseteq V^-$ be the sets minimizing $g()$
among all feasible pairs of sets for the relaxed SM2-multi.  Let $S^{*}$ be an
optimal set minimizing the function $f()$ in the (original) SM2-multi formulation. Then, $2f(S^{*}) \geq f(X'^+ \cup X'^-)$.
\end{thm}
{\bf Proof:}  Let $S^{*+}$ and $S^{*-}$ be the copies of $S^{*}$ in $V^+$ and $V^-$
respectively.  Then,
\vspace{-0.05in}
\[
 \begin{array}{ll}
 2f(S^{*}) &  = f(S^{*+}) + f( S^{*-})
\geq g(X'^+ \cup X'^-) = f(X'^+ )+f( X'^-) \\
  & \geq f(X'^+ \cup X'^-)+f (X'^+ \cap X'^-)\geq   f(X'^+ \cup X'^-).     \end{array}
\]

\vspace{-0.06in}
The first inequality holds since $X'^+ \cup X'^-$ is an optimal solution to the
relaxed SM2-multi.  The second inequality follows from the submodularity of the
function $f$. $f(X'^+ \cup X'^-)$ is the value of our solution where an element
is included if either one of its two copies is in $X'^+$ or in $ X'^-$.
\qed

%
%

\noindent {\bf $2$-approximation for round-up SM2-multi}
Theorem \ref{thm:approx} leads immediately to the $2$-approximation result for round-up SM2-multi:

\begin{lemma} \label{lem:approx1}
For round-up SM2-multi and a {\bf general} non-negative submodular objective function $f()$
there is a polynomial time $2$-approximation algorithm.
\end{lemma}
{\bf Proof:}
If both variables $x^+_j , x'_j $ are of value $1$, then we
set the value of $x_j=1$.  Let $V^1=\{ j\in V|x^+_j = x'_j=1\} $ be the set of such variables. If both
$x^+_j , x^-_j $ are of value $0$, we let $x_j=0$, and $V^0=\{ j\in V|x^+_j = x'_j=0\}$ is the set of
these variables. The set of remaining variables, which have exactly one of $x^+_j
, x'_j $ equal to $1$ and the other one equal to $0$, is called $V^{\half}$.

For round-up SM2-multi, the rounded solution is the set $V^1 \cup V^{\half} =X'^+ \cup
X'^-$.  From Theorem \ref{thm:approx} this is a $2$-approximate solution for SM2-multi.
\qed

We prove next the approximation result for SM problems without the round up property.
To do that
we first demonstrate that a SM2-multi is equivalent to a respective SM-MIN-2SAT problem:

\begin{thm}\label{thm:2SAT1}
The set of SM2-multi constraints is equivalent to the constraints of SM-MIN-2SAT on
at most $nU$ binary variables and $O(m U ) $ constraints, for $U=\max _{i\in V} u_i$, in that both have the same sets of feasible solutions.
\end{thm}
{\bf Proof:}
For a general
constraint of the form, $a_{ki} x_i\ +\  a_{kj} x_j\  \ge\  c_{k} $, consider
the case where both $a_{ki}$ and $a_{kj}$ are positive, and assume without loss of generality that
$0<c_{k}< a_{ki}u_i + a_{kj}u_j $.  The other cases where one coefficient is negative
(and the constraint is monotone), or both are negative, are similarly
``binarized".

For every $\ell$ ($\ell =0,\ldots ,u_i$), let $ \alpha_{k \ell } = \left\lceil
{c_k - \ell a_{{ki}}  \over a_{{kj}}} \right\rceil -1\ . $ For any integer
solution $\x$, $a_{{ki}} x_i + a_{{kj}} x_j  \ge c_k$ if and only if for every
$ \ell$ ($ \ell =0,\ldots ,u_i-1$),
$ \mbox{either }\  x_i > \ell\ \mbox{ or }\ x_j > \alpha_{k \ell},$
or, equivalently,
$
\mbox{either }\  x_i \ge \ell +1\ \mbox{ or }\  x_j \ge \alpha_{k \ell} +1\ ,$
which can be written as  $x_{i,\ell +1} + x_{j, \alpha_{k\ell}+1 } \ge 1\ .$

Obviously, if $ \alpha_{k\ell} \ge u_j$, then we fix the variable, $x_{i,\ell
+1} = 0$.

If the above transformation is applied to a monotone system of inequalities,
then the resulting 2-SAT integer program is also monotone. More precisely, for a constraint of the form $a_{{ki}} x_i - a_{{kj}} x_j  \ge c_k$ the set of binarized constraints are all of the form $x_{ip}\geq x_{jq}$, or the reverse inequality, for some values of $p$ and $q$. To see that, note that if $x_j\geq \ell$ then $x_i \geq \left\lceil
{c_k + \ell a_{{kj}}  \over a_{{ki}}} \right\rceil =\beta _{k\ell}$.  For this condition to be satisfied $x_{j\ell}\leq x_{i,\beta _{k\ell}}$, which is a monotone, closure, constraint.

Altogether we have replaced one original constraint on $x_i$ and $x_j$ by at most $u_i+1$
constraints on the variables $x_{i \ell }$ and $x_{j \ell  }$. The other cases,
corresponding to different sign combinations of $a_{{ki}}$, $a_{{kj}}$, and
$c_k$, can be handled in a similar way.  This completes the proof of Theorem \ref{thm:2SAT1}.
\qed

With Theorem \ref{thm:2SAT1}, it is sufficient to show the respective result For (SM-MIN-2SAT).
It is shown next that the rounded (up or down)
solution is $Z\cup V^1$, for $Z=\{ i| z_i=1, i\in V^{\half} \}$ and $Z\subseteq V^{\half} $. In contrast to round-up SM2-multi, here $Z$ may be a strict subset of $Z\subset V^{\half} $.

\begin{lemma}\label{lemma-non-cover}
For a submodular monotone function $f()$, any feasible rounding (up or down) of the variables in $V^{\half}$ yields
a $2$-approximate solution for (SM2).
\end{lemma}
{\bf Proof:} Since $V^1 \cup V^{\half} =X'^+ \cup
X'^-$ and $Z\subseteq V^{\half} $, it follows from the monotonicity of function
$f$ that, $f(V^1 \cup Z)  \leq f(X'^+ \cup X'^-)$.
 From Theorem \ref{thm:approx} we conclude that $ f(V^1 \cup Z)   \leq 2f(S^*)$ thus demonstrating a polynomial
time $2$-approximation algorithm for any SM2 optimization of a monotone submodular function. \qed

It remains to show that if there exists a feasible solution to SM2-multi then it is possible to find a rounding of the variables in $V^{\half}$ that yield a feasible solution to SM2-multi.  Furthermore, such feasible solution can be found in polynomial time.

Assume that the set of SM2 constraints has a feasible integer solution denoted by $z_1,\ldots,z_n$.
(This can be tested in polynomial time by finding a feasible solution to the respective 2SAT problem using the linear time algorithm of \cite{EIS}.)

Let the optimal solution to the monotonized constraints problem be $m_i^+$ and
$m_i^-$.  The first quantity, $m_i^+$, is the number of variables $x_i^{(p)+}$ that belong to $X^+$.  Recall that because of the constraints $x_i^{(p)+}\geq x_i^{(p+1)+}$, there will be a consecutive sequence of variables $x_i^{(p)+}$ that are equal to $1$, for $p=0,1,\ldots ,q_i$ followed by a sequence of $0$s.  Hence, $m_i^+ =q_i$.  Similarly, $m_i^- =q'_i$ is the largest index $p$ such that $x_i^{(p)-}=1$.

For $i=1,\ldots,n$, let $m_i^* = \frac{1}{2}(m_i^+  -  m_i^- )$.
We define the following solution vector, denoted by
$\l=(\ell_1,\ldots,\ell_n)$, where for $i=1,\ldots,n$:
\[\ell _i= \left\{ \begin{array}{ll}
 \min\{m_i^{+},-m_i^{-}\} & \mbox{if} \  z_i\le \min\{m_i^{+},-m_i^{-}
\},\\
 z_i & \mbox{if} \
\min\{m_i^{+},- m_i^{-}\} \le z_i \le \max\{m_i^{+},-m_i^{-}\},\\
\max\{m_i^{+},-m_i^{-}\} & \mbox{if} \
z_i \ge \max\{m_i^{+},-m_i^{-} \}\ .
                \end{array}
\right. \]

We now prove that the vector $\l$ is feasible:

\begin{lemma}
The vector $\l$ is a feasible solution to the given SM2-multi problem. %
\end{lemma}
{\bf Proof:}
Let $ax_i + bx_j \ge c$ be an inequality where $a$ and $b$ are
nonnegative. We check all possible cases.
If $\ell_i$ is equal to $z_i$ or $\min \{m_i^+, -m_i^-\}$,
and $\ell_j$ is equal to $z_j$ or
$\min \{m_j^+ , -m_j^- \}$,
then clearly,
$
a\ell_i +b\ell_j \ge az_i +bz_j \ge  c\ .
$
Suppose $\ell_i \ge z_i$ and $\ell_j =
\max\{m_j^+, -m_j^- \}$.
By construction, we know that
$ a m_i^+ -bm_j^- \ge c\ \mbox{ and }\ -a m_i^- +bm_j^+\ge c\ .$\\
If $\ell_i \ge -m_i^-$, then,
$
 a\ell_i +b\ell_j \ge -am_i^- +bm_j^+ \ge  c\
$.
Otherwise,
$
a\ell_i +b\ell_j \ge am_i^+ -bm_j^- \ge  c\ .$\\
The last case is when $\ell_i =
\max\{m_i^+, -m_i^- \}$, and
$\ell_j =
\max\{m_j^+ , -m_j^- \}$.  In this case,\\
$
a\ell_i +b\ell_j \ge\  am_i^+ -bm_j^- \ge  c\ .
$
The other types of inequalities are handled similarly.
\qed

 The feasibility of the vector $\l$ for the set of constraints implies that,

 \begin{cor} There exists a "rounding" to a set $Z$ satisfying
$$ X^+ \cap X' \subseteq Z \subseteq X^+ \cup X'.$$
\end{cor}

\begin{cor}
If $m_i^+ =m_i^-$ then $z_i=m_i^+ =m_i^-$.
\end{cor}

\section{Submodular minimization over totally unimodular constraints}\label{sec:VCbip}
\subsection{The SM-vertex cover on bipartite graphs is NP-hard}
Iwata and Nagano prove that Switching Submodular Function Minimization (SSFM) is NP-hard, \cite{IN}, even for $f$ strictly monotone.  Let $V$ and $V'$ both consist of $n$ elements, with element $i\in V$ corresponding to element $i'\in V'$, and a subset $X\subseteq V$ corresponding to $X'\subseteq V'$. Let a submodular function $f$ be defined on subsets of $V\cup V'$. The problem SSFM is to find a bi-partition of $V$, $(X\cup Y)$ that minimizes $f(X \cup Y')$.  The proof that the SM vertex cover on bipartite graph is NP-hard is by reduction from SSFM, \footnote{This is to thank Asaf Levin for useful discussions, and in particular for devising this reduction.}:
Given an instance of SSFM with a strictly monotone submodular function $f$.  Construct a bipartite graph on the sets of nodes $V$ and $V'$, with one edge between each  $i\in V$ and $i'\in V'$.  An optimal vertex cover in this graph includes exactly one of each pair of $i$ and $i'$.  The set selected in $V$ is $X$ and its complement- the set selected in $V'$--is $Y$ that give the optimal value of $f(X \cup Y')$.  This reduction is obviously approximation preserving.

Since the constraint matrix of bipartite vertex cover is totally unimodular, we conclude that,
\begin{theorem}\label{thm:SM2-TU-hard} SM minimization over a totally unimodular constraints matrix is NP-hard.
\end{theorem}

\vspace{-0.06in}
\subsection{The bi-submodular vertex cover on bipartite graph}\label{sec:IS}
To generate the intuition for the relationship between bi-submodular vertex cover and SM-closure consider first the vertex cover and closure problems on a bipartite graph $G=(V_1\cup V_2,E)$. We replace the set of edges $E$ by a set of arcs directed from $V_1$ to $V_2$.  Given a feasible closed set $S$ (w.r.t. successors), then $(V_1\setminus S) \cup (V_2 \cap S)$ is a feasible vertex cover. Vice versa, given a feasible vertex cover $D$, then $( V_1\setminus D)\cup (V_2\cap D)$ is a closed set (w.r.t. successors) in $G$.

For a bipartite graph $G=(V_1\cup V_2,E)$ the bi-submodular vertex cover is to minimize $f(D)=f_1(D\cap V_1)+f_2(D\cap V_2)$ for $D$ a vertex cover. Since $D$ is a vertex cover, then  $( V_1\setminus D)\cup (V_2\cap D)$ is a closed set in $G$.  Since $f_1$ is submodular, then $f'(D)=f_1(V_1\setminus D)$ is submodular as well, and so is $g(D) =f'(D)+f_2(D)$.  Since the minimum submodular closure problem $\min _{D\subseteq V_1\cup V_2} g(D)$ is solved in strongly polynomial time, then so is the bi-submodular vertex cover.  The analogous argument proves that the bi-supermodular independent set problem is also solved in strongly polynomial time.


\vspace{-0.06in}
\section{Conclusions}\label{sec:conclusion}
We demonstrate here a unified technique to generate best possible $2$-approximation algorithms for a large
family of constrained submodular optimization with two variables per inequality that are NP-hard.   We introduce here a new
submodular optimization problem -- the submodular-closure problem which is the
building block and major subroutine of the technique.  The results hold also for
submodular minimization over multi-sets. This settles, for the first
time, the approximation and complexity status of a number of submodular
minimization problems including: SM-2SAT; SM-min satisfiability; SM-edge
deletion for clique and SM-node deletion for biclique, as well as establishes that a problem such as the bi-supermodular independent set on bipartite graph is polynomial time solvable, whereas for supermodular objective it is NP-hard.



\end{document}